\documentclass[prl,twocolumn,showpacs,preprintnumbers,amsmath,amssymb,nofootinbib]{revtex4}
\usepackage{subfigure}
\usepackage{graphicx}
\usepackage{bm}

\newcommand{\rd}{\textrm{d}}
\newcommand{\ee}{\begin{equation}}
  \newcommand{\eee}{\end{equation}}
\newcommand{\ea}{\begin{eqnarray}}
  \newcommand{\eea}{\end{eqnarray}}
\newcommand{\phidot}{\dot{\bar\varphi}}
\newcommand{\af}{\alpha}

\newcommand{\afpp}{\left(\frac{\partial \ln \af}{\partial \varphi}\right)_{|\varphi=\varphi_0}}
\newcommand{\ra}{\frac{\Delta\alpha}{\alpha}}
\newcommand{\rarec}{\frac{\Delta\alpha_{\rm rec}}{\alpha}}
\newcommand{\rarecflat}{\Delta\alpha_{\rm rec}/\alpha}
\newcommand{\seta}{Set \rm I}
\newcommand{\setb}{Set \rm II}

\preprint{HD-THEP-04-52}
\begin{document}

\author{Michael Doran}
\affiliation{Institut f\"ur Theoretische Physik, Universit\"at Heidelberg,Philosophenweg 16, 69120 Heidelberg, Germany}
\affiliation{Dept. of Physics \& Astronomy, Dartmouth College, 6127 Wilder Laboratory, Hanover, NH 03755}
\title{Can we test Dark Energy with Running Fundamental Constants ?}
\begin{abstract}
We investigate a link between the running of the fine structure constant $\alpha$
and a time evolving scalar dark energy field. Employing a versatile parameterization
for the equation of state, we exhaustively cover the space of dark energy
models. Under the assumption that the change in $\alpha$ 
is to first order 
given by the evolution of the Quintessence field, we show that current Oklo, Quasi Stellar
Objects and 
Equivalence Principle observations restrict the model parameters considerably
stronger than observations of the Cosmic Microwave Background, Large Scale
Structure and Supernovae Ia combined. 
\end{abstract}

\maketitle
Observations of Supernovae Ia (SNe Ia) \cite{Riess:2004nr}, the Cosmic
Microwave Background (CMB) \cite{Spergel:2003cb,Readhead:2004gy,Goldstein:2002gf,Rebolo:2004vp} 
and Large Scale
Structure (LSS) \cite{Tegmark:2003ud,Hawkins:2002sg} all point towards
some form of dark energy. Over the years, theorists have come up with various models to explain
the nature of dark energy
\cite{Wetterich:fm,Ratra:1987rm,Caldwell:1997ii,Caldwell:1999ew,Freese:2002sq,Bento:2002ps,Bousso:2000xa}
(to name a few).
In particular, it seems plausible that dark energy may be described by
an (effective \cite{Kolda:1999wq,Peccei:2000rz,Doran:2002bc,Carroll:2003st}) scalar field. 
Provided no symmetry cancels it, there
will be a term in the effective Lagrangian coupling baryonic matter to
the scalar field. If this field evolved over cosmological times,
such a coupling would lead to a time dependence of the coupling ``constants''
of baryonic matter \cite{Wetterich:1987fk,Sandvik:2001rv,Damour:2002nv,Wetterich:2002wm,Wetterich:2003jt} . 
Indeed, bounds on the time variation of 
these ``constants'' restrict the evolution of the scalar field and the 
strength of this coupling \cite{Wetterich:1987fk}.

Since the days of Dirac \cite{Dirac:1937ti}, it has been speculated
that the fundamental constants of nature may vary.  In a realistic GUT
scenario, the variation of different couplings is interconnected
\cite{Langacker:2001td,Calmet:2001nu,Mueller:2004gu}.  We will ignore this interdependence in the
following and concentrate on a change in the fine structure
constant $\af$ with all other ``constants'' fixed. Lacking 
detailed knowledge of the dependence of $\alpha$ on the scalar field, we 
make the {\it Ansatz} \footnote{This form has previously been chosen in 
\cite{Chiba:2001er,Anchordoqui:2003ij,Lee:2003bg}; 
Clearly, an expansion to higher order or an analytic dependence is also conceivable
\cite{Parkinson:2003kf} and may fit the data better.}
\ee
\af(\varphi)  = 
 \af_0 + \af_0  \afpp \left [\varphi(z) - \varphi_0  \right].   \label{eqn::alpha}
\eee
Thus, fixing the Taylor coefficient $\partial \ln \alpha / \partial \varphi$ determines the evolution of $\alpha$ 
as a function of $\varphi$. 
This has immediate consequences: 
if the fine structure was indeed different at higher redshifts, the scalar
field must have evolved since then.  Yet, the Oklo nuclear reactor strongly limits the change
of $\alpha$ at low redshifts $z\approx 0.1$. In our {\it Ansatz}, freezing $\alpha$ 
is equivalent to slowing down the evolution of
$\varphi$. As the kinetic energy (of a canonical scalar
field) is given by $T = \frac{1}{2}a^{-2} \phidot^2$, this leads to a drop in $T$ and therefore
to an equation of state $w$ that approaches that of a cosmological constant
\ee
w =  \frac{T - V}{T +V} \to -1.
\eee
In such a scenario, the equation of state
of the scalar field  crosses over from a more positive value 
at earlier times to a behavior that today mimics a  
cosmological constant \cite{Hebecker:2000zb,Caldwell:2003vp,Jassal:2004ej}.
%
%
\begin{figure}[!t]
\includegraphics[height=0.5\textwidth,angle=-90]{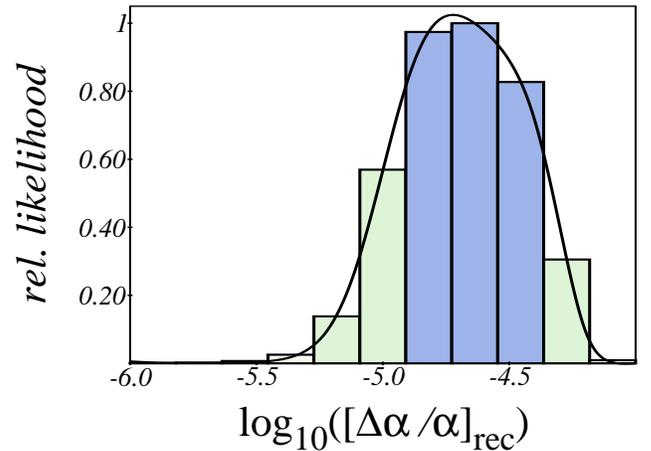}
\caption{\label{fig::rarec}Marginalized likelihood distribution for 
the relative change of the fine structure constant quoted as 
$\log_{10}(\rarecflat)$, so that $\rarecflat = - 10^{\log_{10}\left(\rarecflat\right)}$.
The distribution is obtained from \seta, i.e.
Equivalence Principle, Oklo and QSO observations which are all at considerably lower redshifts
than recombination.
The dark shaded (blue), moderately shaded (green)
and light (red) rectangles correspond to one, two and three $\sigma$
confidence regions. The distribution peaks at $-4.7$ and has width $+0.36, -0.26$.
Thus the relative change of $\alpha$ at recombination is $\rarecflat = -2^{+0.9}_{-2.3} \times 10^{-5}$.}
\end{figure}
%
%
%
\begin{figure}[!t]
\begin{center}
\includegraphics[width=0.45\textwidth]{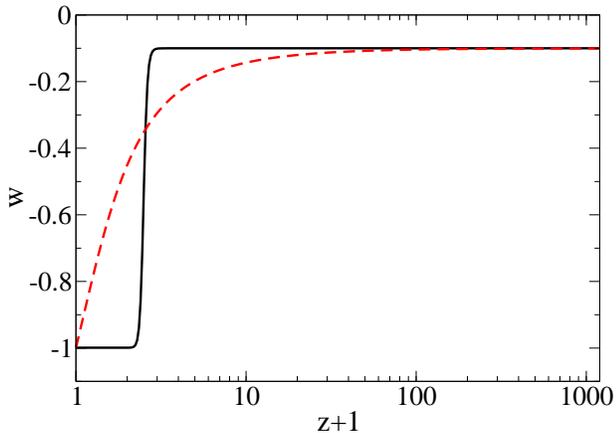}
\caption{Equation of state for  two different parameter sets of Equation \eqref{eqn::pier} versus redshift. The
equation of state in the early Universe is given by $w_m$ (here it is
set to $-0.1$), while the one today is given by $w_0$ (here
$-0.99$). The scale factor of the cross-over is determined by $a_c$
($0.3$ for the solid curve, $0.7$ for the dashed curve), while the
width of the transition is adjusted by $\Delta$ ($0.01$ for the solid
curve, $0.1$ for the dashed one).}
\label{fig::w}
\end{center}
\end{figure}
In order to describe such models, we employ the parameterization
of \cite{Corasaniti:2002vg} 
\begin{multline}\label{eqn::pier}
w(a) = w_0 + (w_m - w_0) \times \frac{1 + \exp(a_c / \Delta) }{1 + \exp( [a_c-a]/\Delta )} \\
\times \frac{1 - \exp( [1-a]/\Delta)}{1 - \exp(1/\Delta)}.
\end{multline}
This versatile parameterization is characterized by the equation of state of dark energy today $w_0$,
the dark energy equation of state during earlier epochs $w_m$, a cross-over scale factor $a_c$
and a parameter $\Delta$ controlling the rapidity of this cross-over (see also Figure \ref{fig::w} for
illustration).
It has been shown (and we will confirm this later)
that some of the four parameters $w_0,\ w_m, a_c,\ \Delta$ are not too
well constrained by present SNe Ia, CMB and LSS data
\cite{Corasaniti:2004sz}.
%
%
\begin{figure}
\begin{center}
\includegraphics[width=0.45\textwidth]{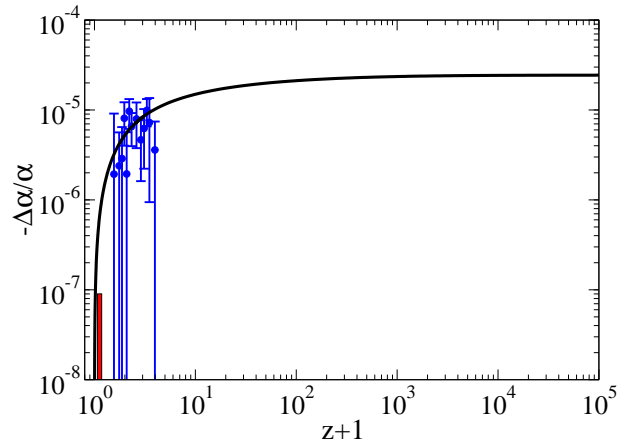}
\caption{Minus the relative change of the fine structure constant 
$\Delta \alpha(z)/\alpha_0 = \left[ \alpha(z) - \alpha_0\right] / \alpha_0$ vs. redshift quoted as $z+1$.
The QSO data is depicted at redshifts $z\sim 2-3$. In addition, we indicate the bound from the
Oklo natural reactor $\Delta \alpha(z\approx 0.13)/\alpha_0 > -9 \times 10^{-8}$ by a 
slim vertical box in the lower left corner. The solid black curve is 
a model we picked at random from the Monte Carlo and which has parameters
 $\Omega_m h^2=0.12$, $h=0.55$,$\ w_0=-0.95$,$\ w_m=-0.26$, $\ a_c=0.8$, $\log_{10}(\Delta)=-2.8$ and
$\rarecflat = -2.4 \times 10^{-5}$. Please note the moderate running of $\alpha$ at early times which is a direct
result of the scalar field dynamics: with $w_m = 0$,  one gets  $\varphi \propto \ln (1+z)$ and hence
a moderate evolution at high redshifts. For $w_m < 0$, the running is suppressed further as the
total energy and hence the available kinetic energy of dark energy is considerably lower than for $w_m=0$.
Hence, the change of $\alpha$ at nucleosynthesis would have negligible effect \cite{Mueller:2004gu}.
}
\label{fig::alpha}
\end{center}
\end{figure}
%
%
In principle, a change of $\alpha$ at high redshifts will alter
recombination. This effect has been discussed in
\cite{Hannestad:1998xp,Kaplinghat:1998ry,Huey:2001ku} and we have modified the {\sc recfast}
\cite{Seager:1999bc} implementation of {\sc Cmbeasy}
\cite{Doran:2003sy} to accommodate this deviation from standard
recombination. Our results agree well with those of
\cite{Kaplinghat:1998ry,Sandvik:priv}.  However, we will soon see that
 Oklo and Quasi Stellar Objects (QSO) observations when explained in our framework limit
the relative change of $\alpha$ at and during recombination.
We will in the following quote this change at the pivotal redshift
$z_{\rm rec}=1100$.
\footnote{We identify $z=1100$ with the redshift of recombination.
This is purely semantics and the mismatch
with the true redshift of recombination is irrelevant.}
Typically, in our scenario 
$\rarecflat \equiv \left[ \alpha(z_{\rm rec}) - \alpha_0\right]/\alpha_0 < 10^{-4}$ (see also Figure \ref{fig::alpha}). 
For such minute changes in $\alpha$, the recombination history remains
practically unaltered. Therefore, considerations like
that of \cite{Huey:2001ku,Rocha:2003gc} have negligible effect.
This fact allows us to split the analysis in two parts (\seta\ and \setb):
\seta\ with Equivalence principle, Oklo and QSO observations for
which 
we do not need to include cosmological perturbations. \setb\
with WMAP, ACBAR, CBI, VSA, SDSS and SNe Ia 
\cite{Spergel:2003cb,Readhead:2004gy,Goldstein:2002gf,Rebolo:2004vp,Tegmark:2003ud,Riess:2004nr} 
observations (and hence cosmological
perturbations), but constant $\alpha=\alpha_0$. In both 
cases, 
we employed the Markov Chain Monte Carlo (MCMC) \cite{Lewis:2002ah}
from the {\sc AnalyzeThis} package \cite{Doran:2003ua} of {\sc Cmbeasy} \cite{Doran:2003sy}. 

For completeness, let us describe the well known reconstruction of 
the scalar field evolution from some parameterization $w(a)$:
the starting point is a reconstruction of $\rho_{de}(a)$ by integrating
$\rd \ln \rho_{de} = -3[1+w(a)] \rd \ln a$. This fixes $\dot\varphi(a)$, because
$T = \frac{1}{2}(1+w) \rho_{de}  = \frac{1}{2} a^{-2} \dot\varphi^2$.
From this, $\varphi(a)$  simply follows by integrating $\dot\varphi$, where we
use an integration constant to fix $\varphi_0 = 0$ for convenience.
Demanding that $\alpha$ should change by $\rarecflat$ at $z_{\rm rec}=1100$ and using 
$\varphi_0=0$,  we can
re-write Equation \eqref{eqn::alpha}  as
\ee
\alpha(z) = \alpha_0 + \alpha_0 \left(\rarec\right) \frac{\varphi(z)}{\varphi_{\rm rec}}.
\eee
\newlength{\threewidth}
\setlength{\threewidth}{0.45\textwidth} 

\newlength{\fourwidth}
\setlength{\fourwidth}{0.41\textwidth} 

\begin{figure*}[!t]
\subfigure[]{
  \includegraphics[height=\threewidth,angle=-90]{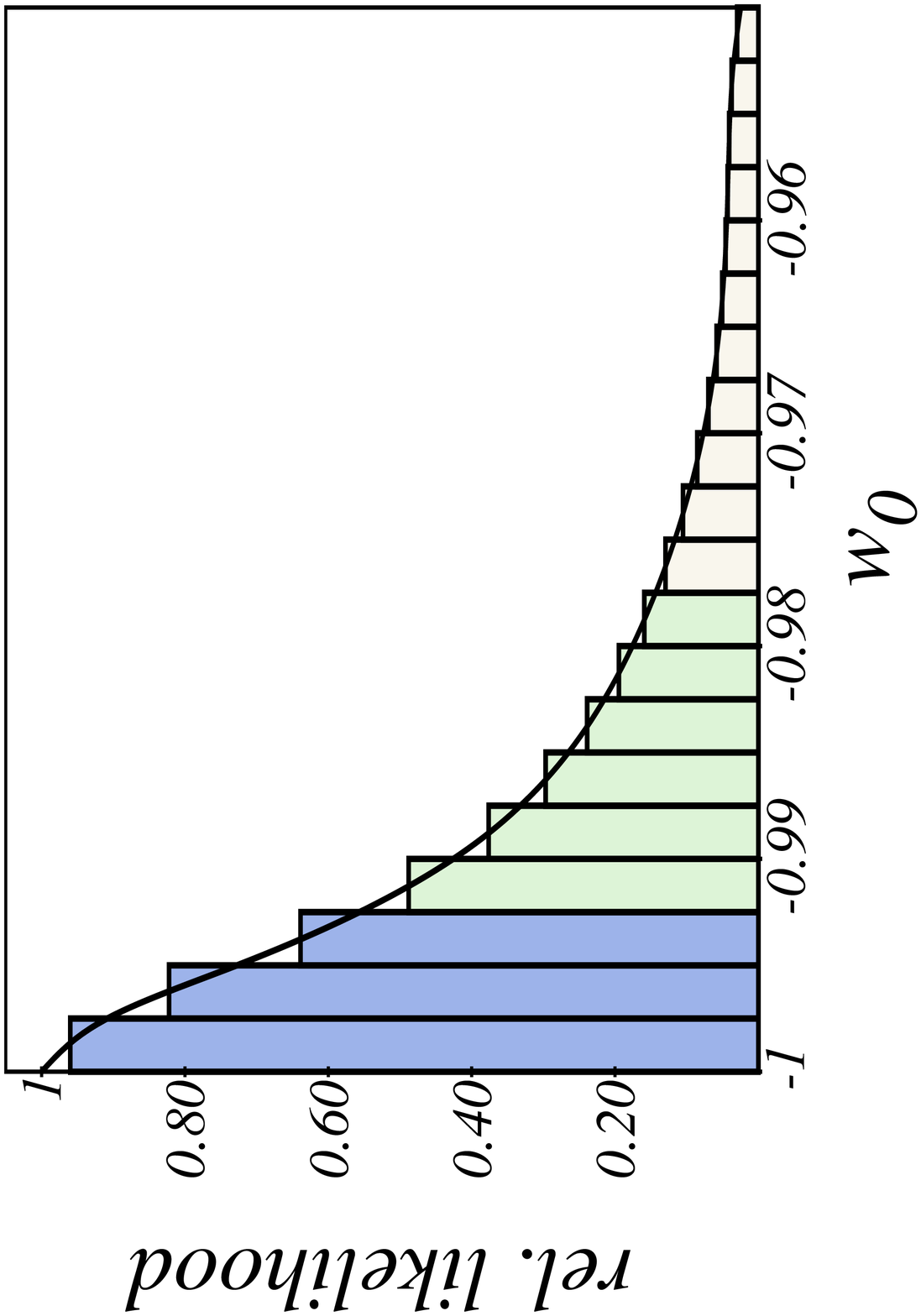}
}
\subfigure[]{
  \includegraphics[height=\threewidth,angle=-90]{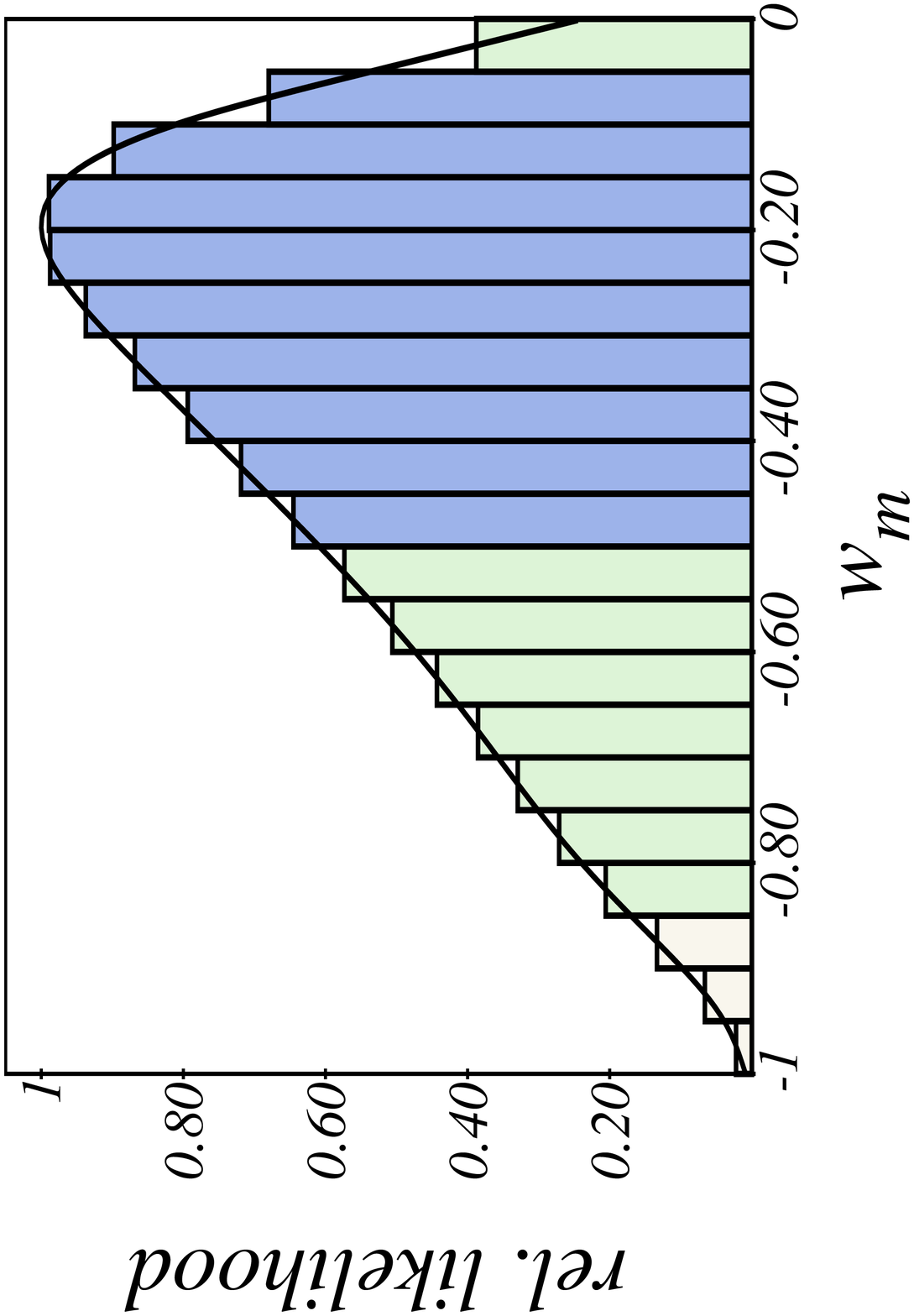}
}
\subfigure[]{
  \includegraphics[height=\threewidth,angle=-90]{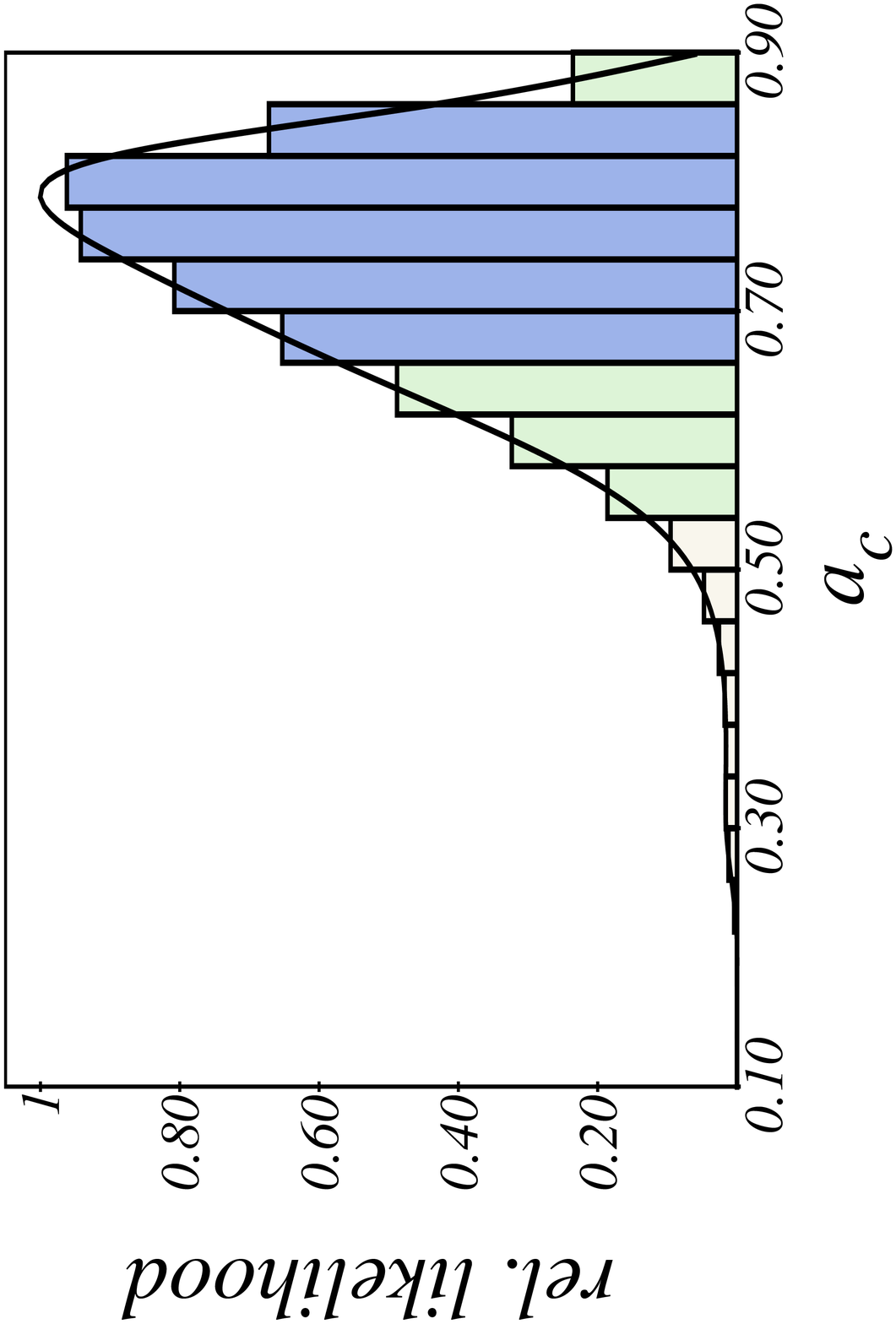}
}
\subfigure[]{
  \includegraphics[height=\threewidth,angle=-90]{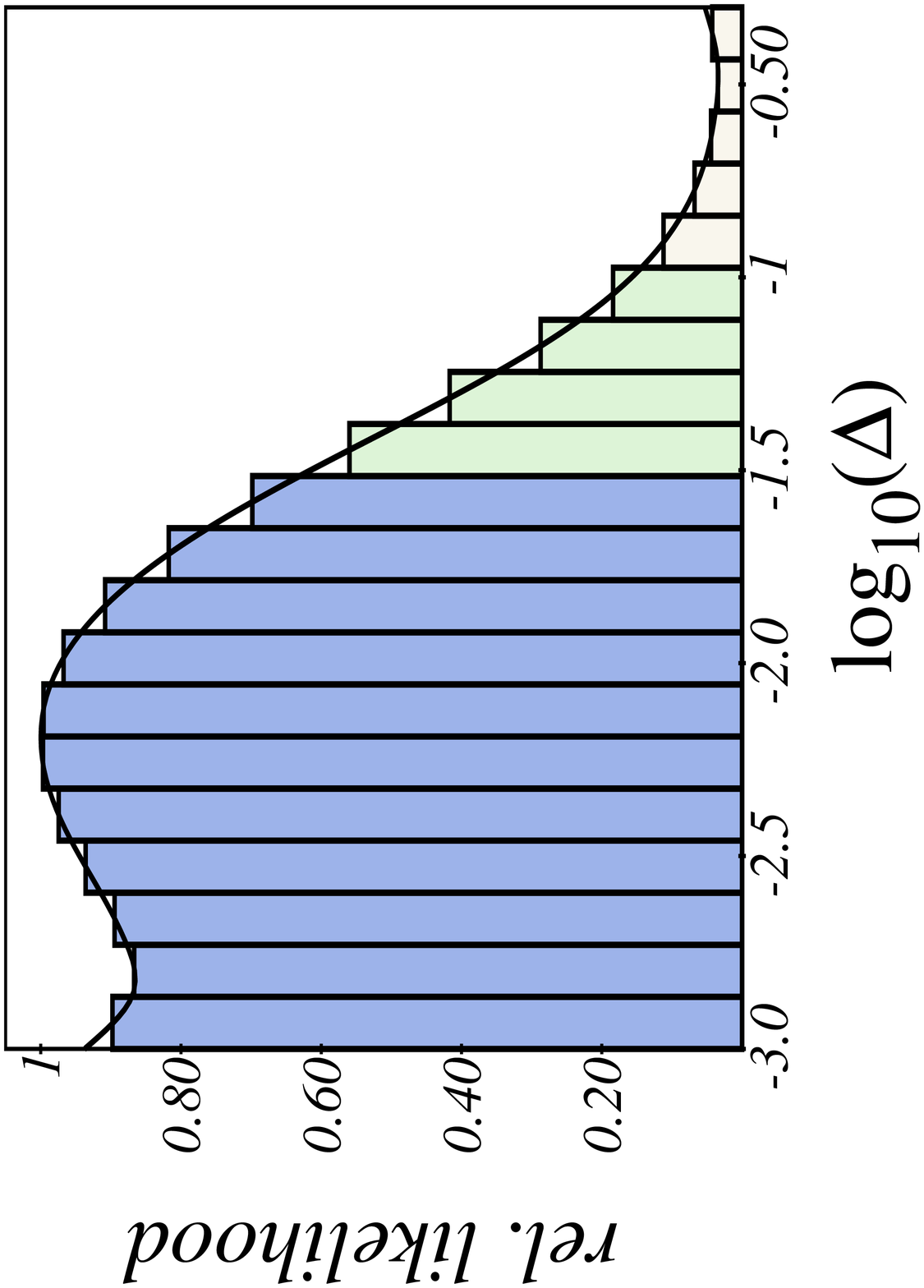}
}
\caption{\label{fig::combined}
Marginalized likelihood distribution from the combination of \seta\ and \setb. 
The dark shaded (blue), moderately shaded (green)
and light (red) rectangles correspond to one, two and three $\sigma$
confidence regions. At $2\sigma$ confidence
level, we get $w_0 < -0.97$ and $\Delta < 0.1$, while with  $1\sigma$ error bars,
we have  $w_m = -0.2^{+0.13}_{-0.30}$ and  $a_c = 0.79 ^{+ 0.06}_{-0.12}$.
}
\end{figure*}

For \seta, 
the parameters of
the model were $\Omega_m h^2$,$\ h$,$\ w_0,\ w_m$,$\ a_c$,$\ \log_{10}(\Delta)$ and
$\log_{10}\left(\rarecflat\right)$.\footnote{We picked negative $\Delta\alpha/\alpha$, i.e. 
$\rarecflat = - 10^{\log_{10}\left(\rarecflat\right)}$.}
Please note that we omit the spectral index $n$, optical depth $\tau$ and baryon
fraction $\Omega_b h^2$ which are present in the \setb-analysis, because \seta\
data does not contain information on these parameters. Thus, the distribution
in $n$,$\ \tau$ and $\Omega_b h^2$ is flat and our simulation for \seta\ can
be thought of as trivially marginalized over $n$,$\ \tau$ and $\Omega_b h^2$. 
The Equivalence Principle is tested via the differential acceleration
$\eta$ between test particles of equal mass but different composition
\cite{Wetterich:2003jt}
\ee
\eta = -3 \times 1.75 \cdot 10^{-2} \left(\frac{\partial \ln \alpha}{\partial \varphi}\right)_{\varphi=\varphi_0}^2 \Delta R(1 + Q).
\eee 
Here, $\Delta R \equiv \Delta Z / (Z+N) \approx 0.1$  is the difference in composition and the factor $(1+Q)$
encodes the theoretical uncertainty in calculating $\eta$. As the true underlying theory is unknown, 
we 'guesstimate' conservatively $(1+Q) \in [10^{-2},10^2]$ which is used as a flat prior. As current experiments
give a null result \cite{baessler}
\ee
\eta = 0 \pm 3\times 10^{-13},
\eee
$(1+Q)=10^{-2}$ (i.e. the lower bound) always agrees best with observations.

Roughly $1.8$ billion years ago (i.e. $z\approx 0.1$), the Oklo natural reactor was up and running. 
Until recently, the reactor was seen as providing yet another null result, namely \cite{Damour:1996zw}
\ee\label{eqn::damour}
\ra_{\rm Oklo} = 0^{+12}_{-9}     \times 10^{-8}.
\eee

Yet, the latest calculation \cite{Lamoreaux:2003ii} yields a positive detection of 
\ee
\ra_{\rm Oklo} = 45^{+7}_{-15}       \times  10^{-8}  
\eee
To complicate matters further \cite{Martins:2004ni}, Fujii et. al \cite{Fujii:2003gu} find two values, 
\ea
\ra_{\rm Oklo} &=& (-0.8 \pm 1.0) \times 10^{-8},\\
\ra_{\rm Oklo}  &=&(8.8 \pm 0.7) \times 10^{-8}.
\eea
We will take a conservative stand here (given the scatter in the
calculations) and use the null-result of \cite{Damour:1996zw}, Equation
\eqref{eqn::damour}.  
\begin{figure*}[!t]
\begin{tabular}{lr}
%
%
   \includegraphics[height=\fourwidth,angle=-90]{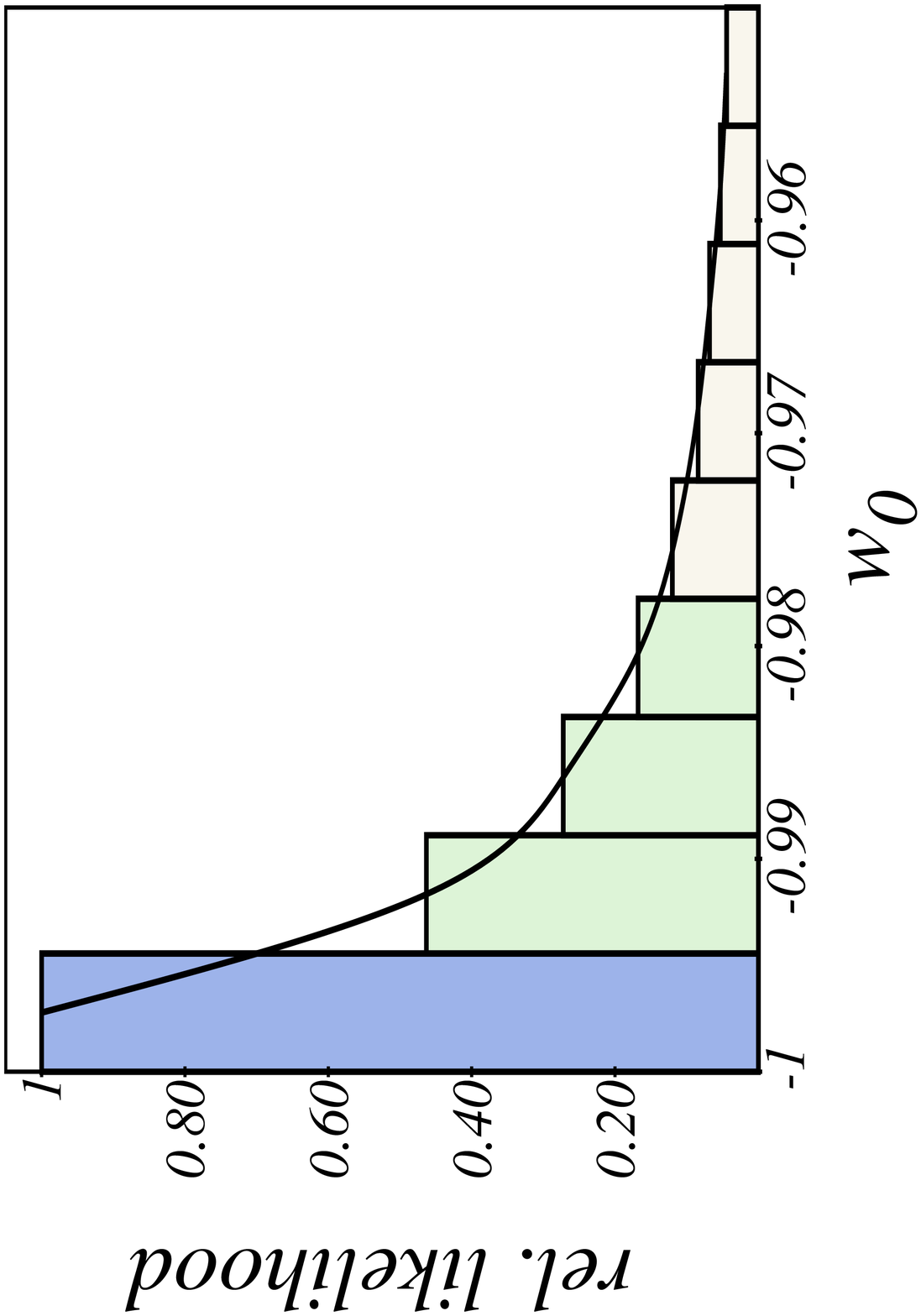} &
  \includegraphics[height=\fourwidth,angle=-90]{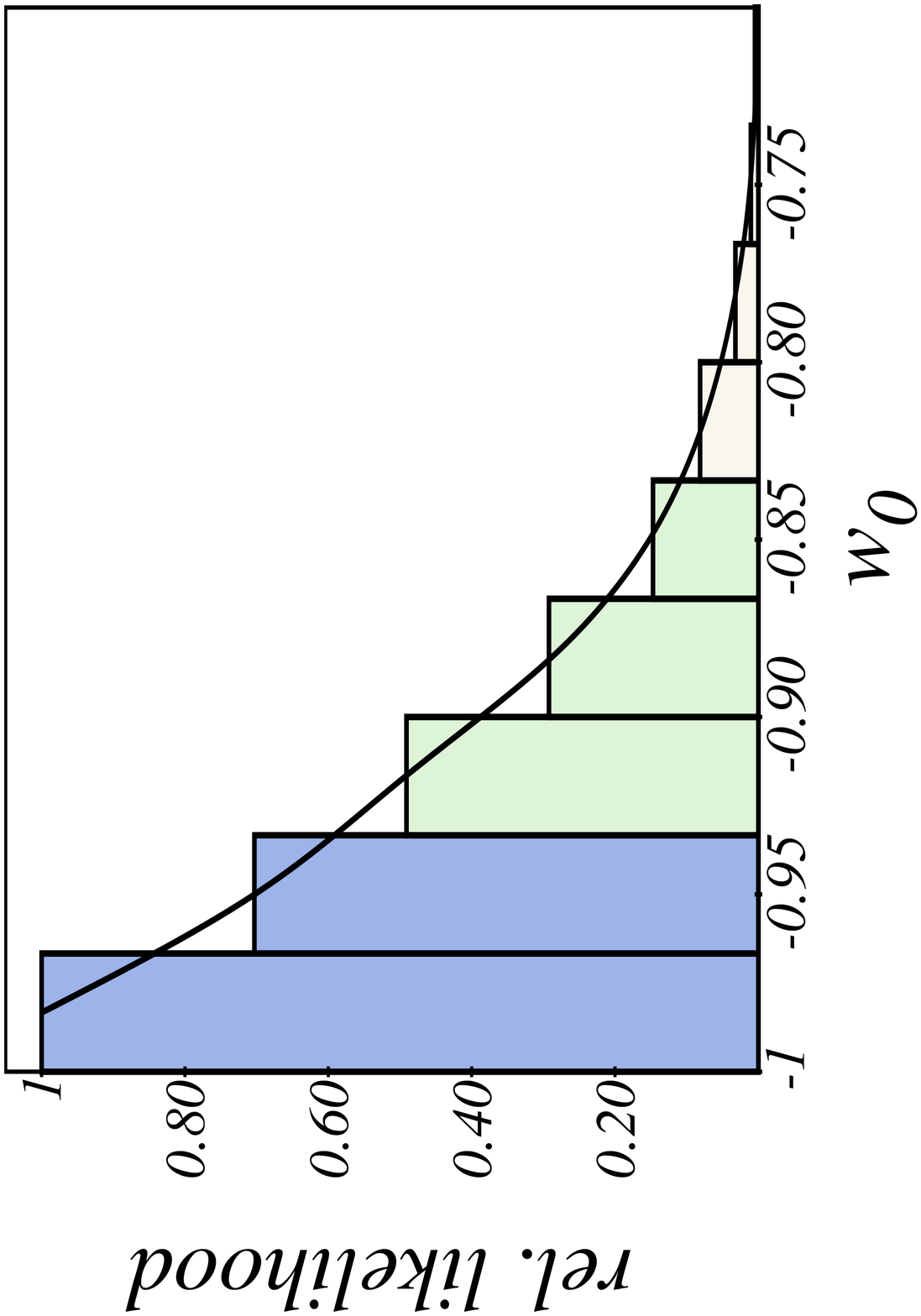}  \\
%
%
 \includegraphics[height=\fourwidth,angle=-90]{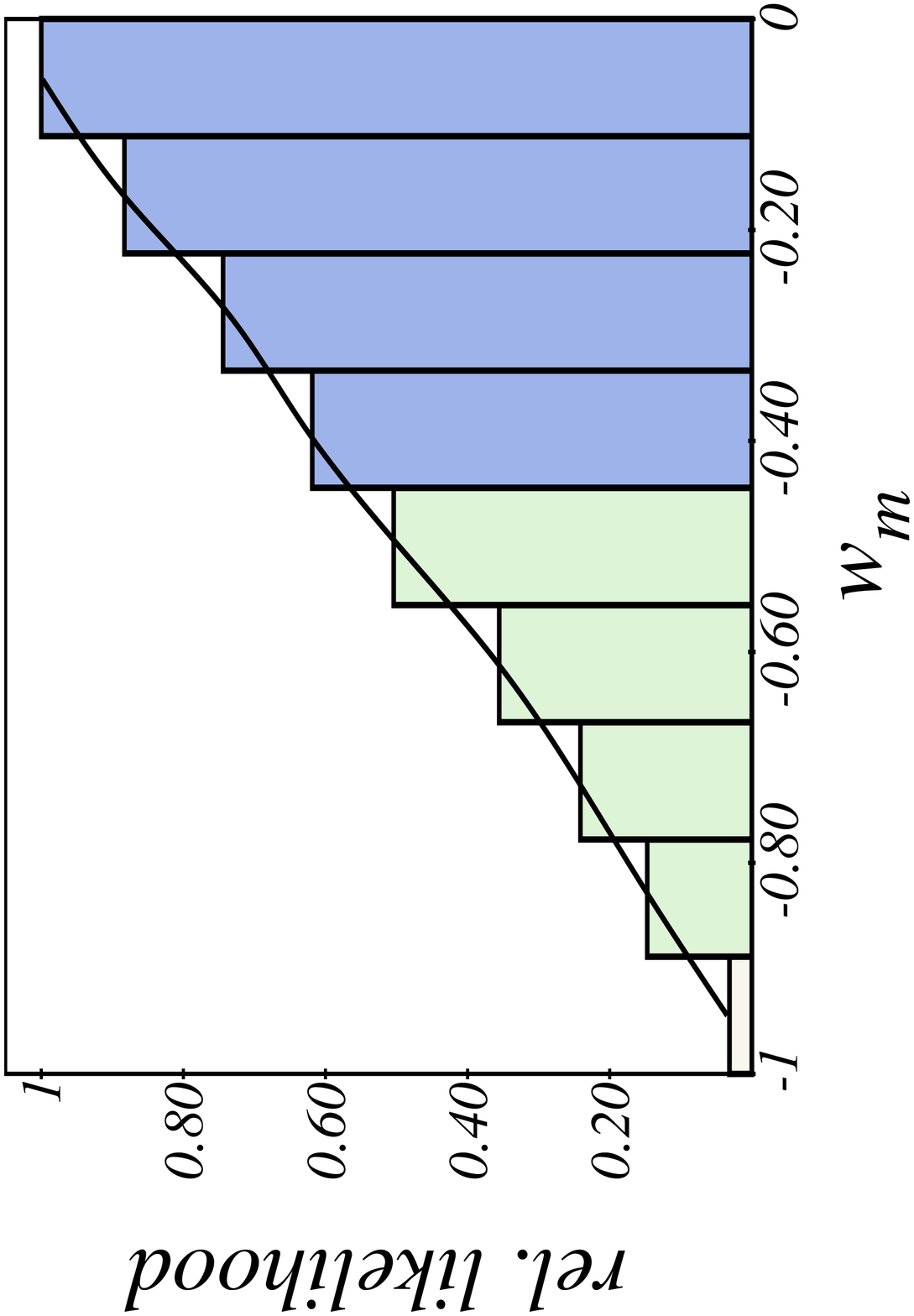} &
 \includegraphics[height=\fourwidth,angle=-90]{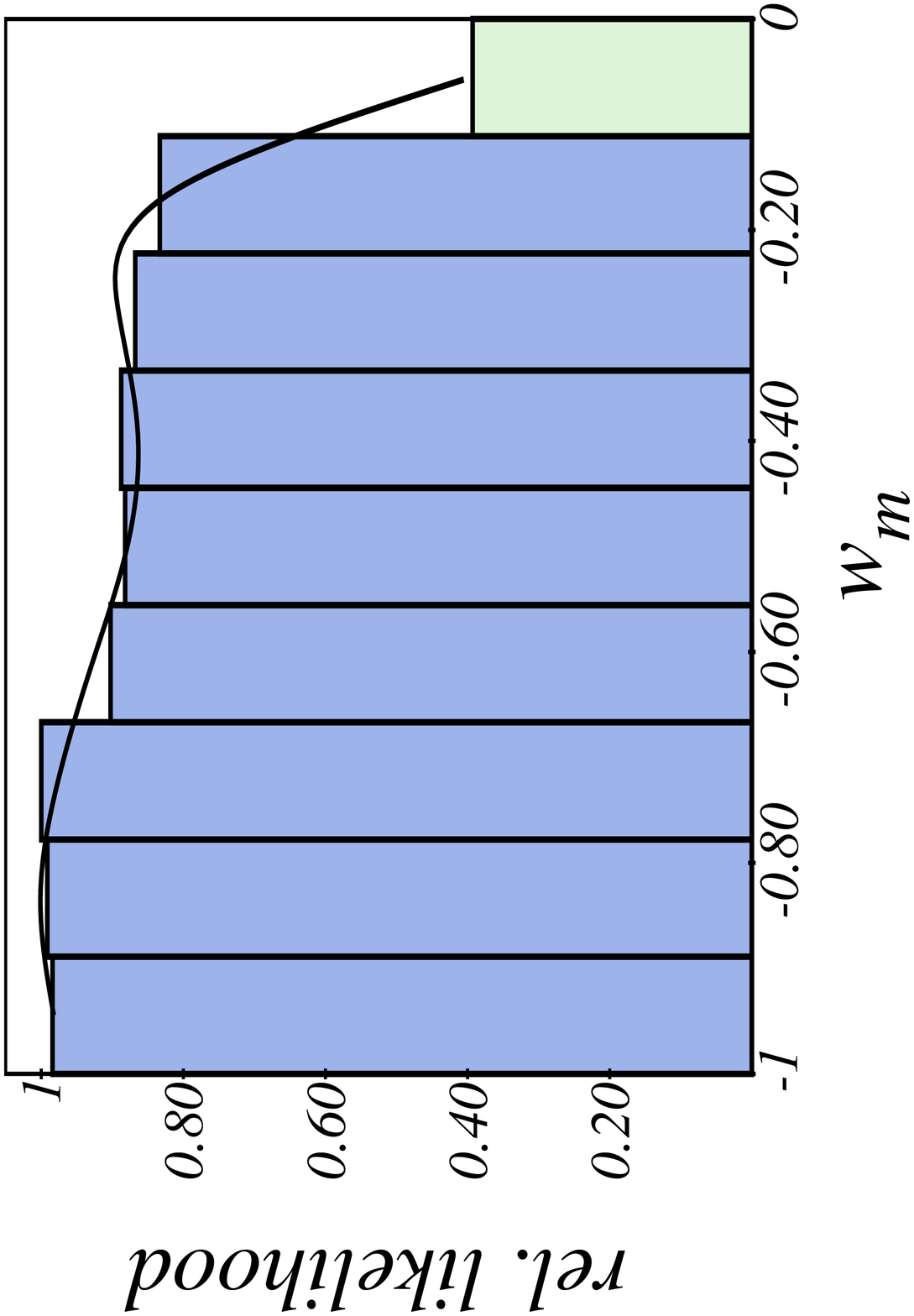} \\
%
%
 \includegraphics[height=\fourwidth,angle=-90]{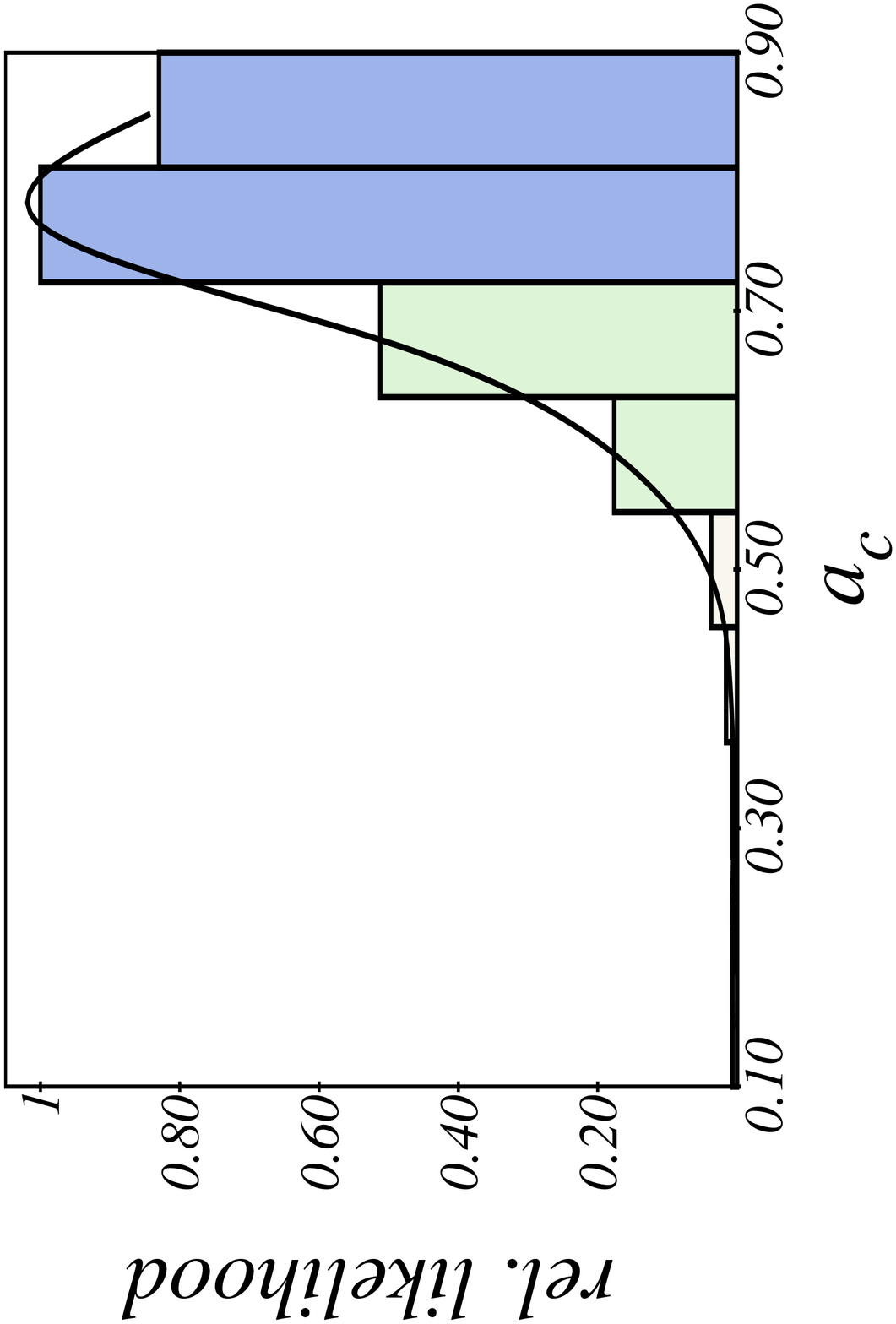} &
 \includegraphics[height=\fourwidth,angle=-90]{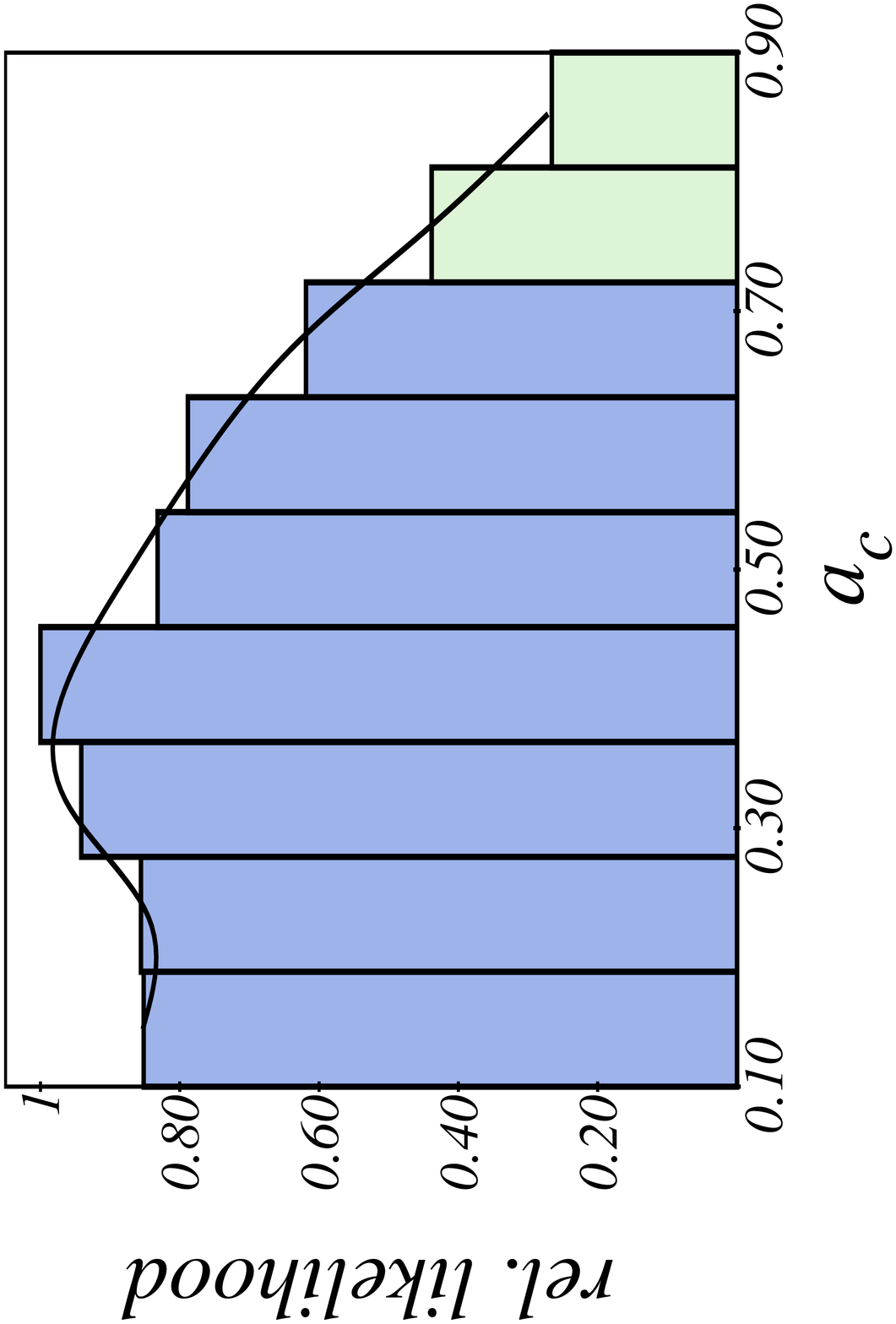} \\
%
%
 \includegraphics[height=\fourwidth,angle=-90]{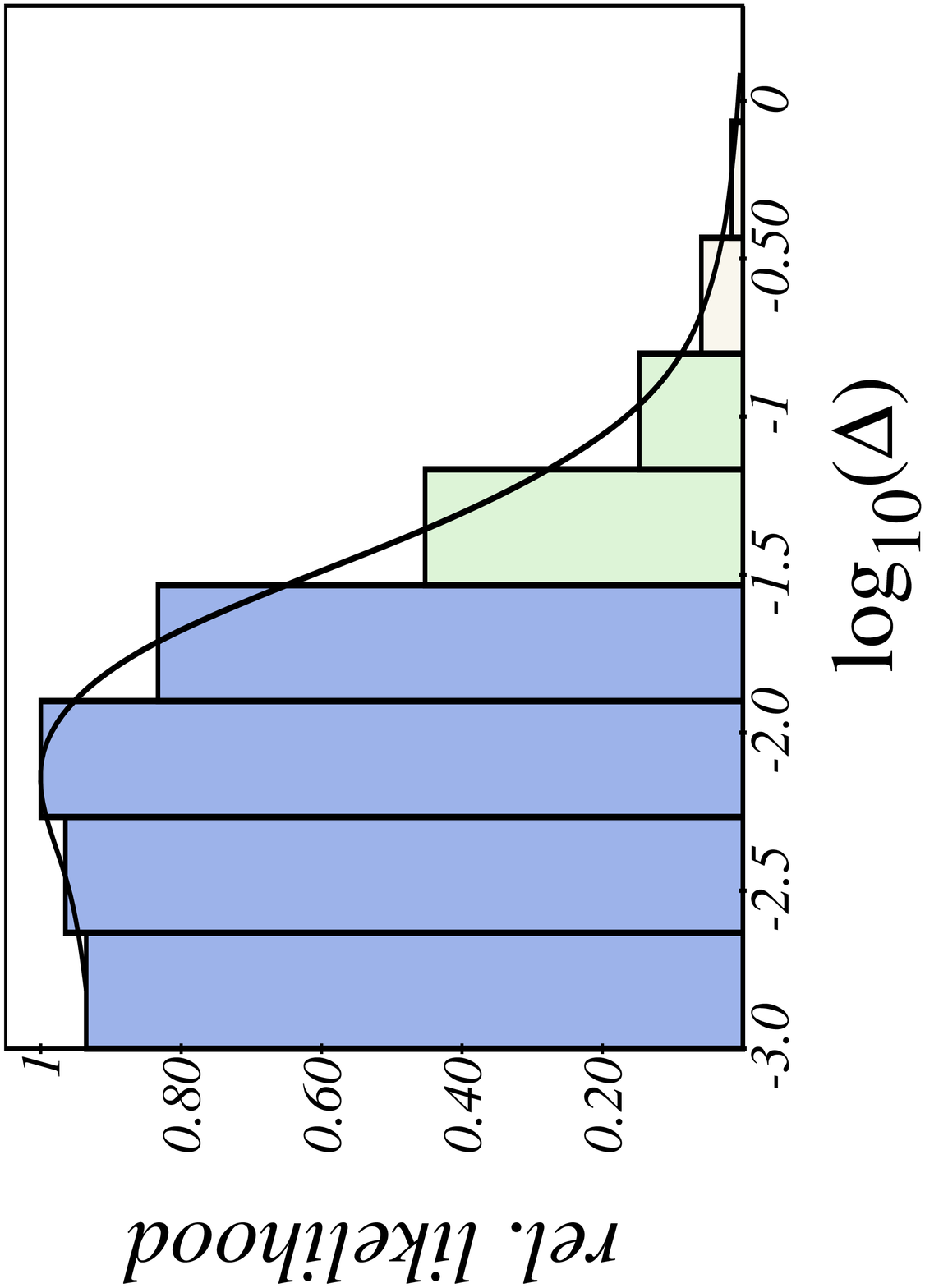} &
 \includegraphics[height=\fourwidth,angle=-90]{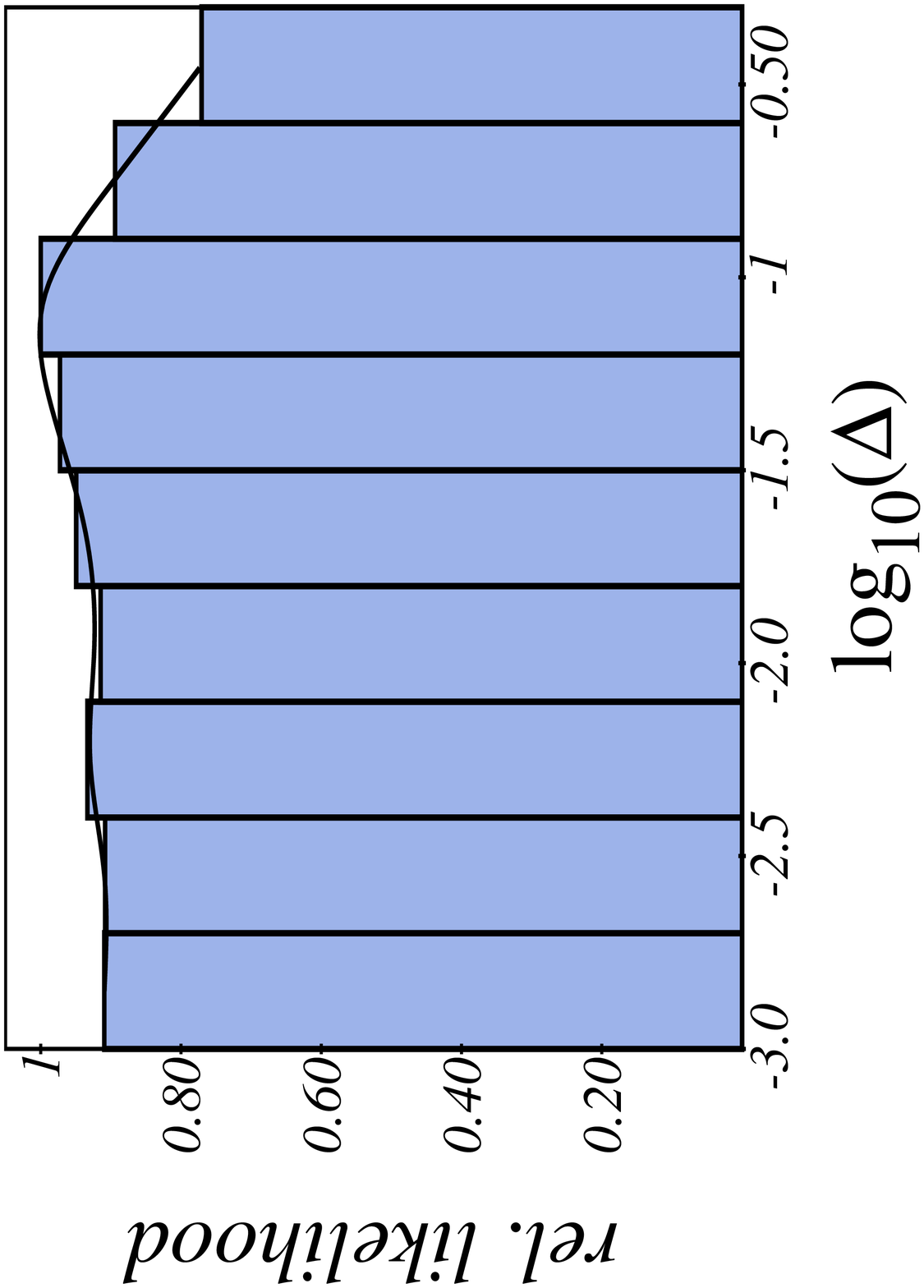} 
\end{tabular}
\caption{\label{fig::pert}
Marginalized likelihood distribution for the parameters of  Equation \eqref{eqn::pier}.
The left column derives from Oklo, QSO and Equivalence Principle observations, while
the right column uses WMAP, ACBAR, CBI, VSA, SDSS and SNe Ia data.
Please note that the bound on all parameters on the left is 
considerably tighter than that in the right column. The dark shaded (blue), moderately shaded (green)
and light (red) rectangles correspond to one, two and three $\sigma$
confidence regions.
}
\end{figure*}

For QSO observations, we use\footnote{To
prevent misunderstandings: we used the binned data shown in Figure
\ref{fig::alpha} } the sample of \cite{Murphy:2003hw,Murphy:priv}
which roughly translates into
\ee\label{eqn::webb}
\ra_{\rm QSO} = (-5  \pm 1) \times 10^{-6}.
\eee
There have been contradicting claims about such 
QSO results  \cite{Chand:2004ct}. While time may tell what 
$\alpha$ really was at $z \approx 2$, we stick with Equation \eqref{eqn::webb}
for the purpose of this letter. So we ask: if $\alpha$ really changed 
in the past, what does it tell us about dark energy?
 
The main results of our \seta\ analysis are summarized 
in Figure \ref{fig::rarec}  and the left column of Figure \ref{fig::pert}. 
From Figure \ref{fig::rarec}, we see that $\rarecflat = -2^{+0.9}_{-2.3} \times 10^{-5}$.
Likewise, in  the left column of  Figure \ref{fig::pert}, we see that these measurements 
lead to strong bounds on the equation of state today. At $2\sigma$ confidence
level, $w_0  < -0.97$. In addition, the transition is restricted to
occur  rather swiftly (as seen from $\log_{10}(\Delta)$) and at rather recent times (as
seen from $a_c$).

Turning to our \setb\  analysis,
the parameters were the amount of matter and baryons $\Omega_{m} h^2$ and  $\Omega_{b} h^2$, 
Hubble constant  $h$, optical depth to re-ionization $\tau$ and 
spectral index $n$ plus four parameters $w_0$,$\ w_m$,$\ a_c$,$\ \log_{10}(\Delta)$
of the dark energy model. As said, we did not need to take a varying $\alpha$ into 
account, because the allowed values of $\rarecflat$ from \seta\
are minute in our framework and will not alter standard recombination.\footnote{We
\emph{did} perform a  Monte Carlo including varying $\alpha$. As expected, nothing changed.}
To test this model, we compared to WMAP, VSA, ACBAR, CBI, SDSS and 
SNe Ia measurements 
\cite{Spergel:2003cb,Readhead:2004gy,Goldstein:2002gf,Rebolo:2004vp,Tegmark:2003ud,Riess:2004nr}.
The versatile (and therefore difficult to constrain) parameterization of Equation \eqref{eqn::pier} has 
been under investigation in \cite{Corasaniti:2004sz}. Our results agree well
with those presented in that paper (compare the right column of our Figure \ref{fig::pert} to Figure 6
of  \cite{Corasaniti:2004sz}). With the exception of $w_0$, the constraints are considerably 
less tight than those inferred from the fine structure data.
Combining \seta\ with \setb\ leads to the likelihood distributions shown in
Figure \ref{fig::combined} and represent our main results. At
$2\sigma$ confidence level, we get $w_0 < -0.97$ and $\Delta < 0.1$.
Quoting the $1\sigma$ error bars, we get $w_m = -0.2^{+0.13}_{-0.30}$
and $a_c = 0.79 ^{+ 0.06}_{-0.12}$.

In this letter, we presented a quantitative analysis of a scenario 
in which the running of the fine structure constant is driven
by a scalar dark energy field. We parameterized the evolution of  the scalar field in a 
versatile manner thus covering nearly all of today's 
quintessence models with $w > -1$.  Together with the {\it Ansatz} that 
the induced running of $\alpha$ is linear in the field we found stringent
constraints for dark energy model building. 
For future work, the analysis may be extended in two directions: 
as the running of couplings in realistic GUT scenarios is interdependent, one 
may investigate a running of $\alpha$ together with a running of the Planck mass.
Secondly, the ${\it Ansatz}$ of a linear dependence of $\alpha$ on the dark energy
field may be promoted to higher order or an analytic function. In this case,
a substantial change of recombination or nucleosynthesis physics is well conceivable.

\noindent {\bf Acknowledgments}\, 
I would like to thank Bruce A. Basset, Robert R. Caldwell, Pier-Stefano Corasaniti, 
Manoj Kaplinghat, Martin Kunz, Christian M. M\"uller,
Havard B. Sandvik, Douglas Scott and Christof Wetterich for helpful discussions.
Part of this work was supported by NSF grant PHY-0099543 at Dartmouth College and
DFG grant We1056/6-3 at Heidelberg.
I would like to thank Caltech for hospitality during the course of part of this investigation.

\end{document}